\documentclass[usenatbib]{mnras}
\input epsf

\usepackage{graphicx}
\usepackage{txfonts}

\usepackage{epsfig}

\title[The Sun in transition?]{The Sun in transition? Persistence of
  near-surface structural changes through Cycle\,24}

\author[R. Howe et al.]{R.~Howe$^{1,2}$, G.~R.~Davies$^{1,2}$,
  W.~J.~Chaplin$^{1,2}$, Y.~Elsworth$^{1,2}$, S.~Basu$^{3}$,
  \newauthor S.~J.~Hale$^{1,2}$,  
W.~H.~Ball$^{1,2}$, R.~W.~Komm$^4$\\
$^{1}$School of Physics and Astronomy,
  University of Birmingham, Birmingham, B15 2TT, United
  Kingdom\\ $^{2}$Stellar Astrophysics Centre (SAC), Department of
  Physics and Astronomy, Aarhus University,\\ Ny Munkegade 120, DK-8000
  Aarhus C, Denmark\\ $^3$Department of Astronomy, Yale University, PO
  Box 208101, New Haven, CT 06520-8101, USA,\\
$^4$National Solar Observatory, Tucson, AZ 85719, USA}

\begin{document}

\maketitle

\begin{abstract}


We examine the frequency shifts in low-degree helioseismic modes from the
Birmingham Solar-Oscillations Network (BiSON) covering the period from 1985\,--\,2016, and compare them with a number of global activity proxies well as 
a latitudinally-resolved magnetic index. As well as looking at frequency shifts in different frequency bands, we look at a parametrization of the shift as a cubic function of frequency. While the shifts in the medium- and high-frequency bands are very well correlated with all of the activity indices (with the best correlation being with the 10.7\,cm radio flux), we confirm earlier findings that there appears to have been a change in the frequency response to activity during solar cycle 23, and the low-frequency shifts are less correlated with activity in the last two cycles than they were in Cycle 22. At the same time, the more recent cycles show a slight increase in their sensitivity to activity levels at medium and higher frequencies, perhaps because a greater proportion of activity is composed of weaker or more ephemeral regions. This lends weight to the speculation that a fundamental change in the nature of the solar dynamo may be in progress.

\end{abstract}

\begin{keywords}

methods: data analysis -- methods: statistical -- Sun: helioseismology

\end{keywords}

\section{INTRODUCTION}
\label{sec:intro}

The current solar activity Cycle\,24 has been significantly weaker
than the previous few cycles 
\citep[e.g., see][]{2015LRSP...12....4H}.
These changes
were signposted by the unusually extended and deep solar minimum at
the boundary of Cycles\,23 and~24.  Very few of the predictions
collated by the Solar Cycle\,24 Prediction Panel 
\citep{2008SoPh..252..209P,2012SoPh..281..507P}
forecast the extent of the minimum or the low levels of activity that
followed.

One must go back around one-hundred years to find cycles that show
levels of activity as low as those observed in Cycle\,24, e.g.,
Cycles\,14 and~15 both provide very good matches in traditional
proxies such as the International Sunspot Number (ISN). Tellingly,
this earlier epoch pre-dates both the modern Grand Maximum period
\emph{and} the satellite era. The wide range of contemporary
observations and data products was therefore not available to
characterize and study the Sun during that era.

Some activity indicators dropped to remarkably low values during the
Cycle\,23/24 minimum (e.g. the geomagnetic $aa$-index and the ISN).
Solar wind turbulence, as captured by measures of interplanetary
scintillation, had been declining since the early part of Cycle~23.
There have been results suggesting a decline -- from the Cycle\,23/24
boundary through the rise of Cycle\,24 -- in the average strength of
magnetic fields in sunspots (\citealt{2012ApJ...757L...8L}; see also 
\citealt{2014ApJ...787...22W})
 and others pointing to a change in the size
distribution of spots between Cycles\,22 and 23 
\citep[e.g., see][]{2012JSWSC...2A..06C,2013ApJ...770...89D}.

Helioseismic studies of the internal solar dynamics showed that
the characteristics of the meridional flow altered between Cycles\,23
and~24  
\citep{2010Sci...327.1350H}.
Changes to the meridional flow
have potential consequences for flux-transport dynamo models.
Differences have also been seen in the east-west zonal flows
\citep[e.g.,][]{2013ApJ...767L..20H}
and in the frequency shifts of globally coherent p
modes, which have been weaker than in preceding cycles 
\citep[e.g., see][]{2012ApJ...758...43B,2015A&A...578A.137S,2015ApJ...812...20T,2015MNRAS.454.4120H}.

\citet{2014ApJ...780....5U} have suggested that it was actually a weak
Cycle\,23 that was responsible for the following, extended minimum and
weak Cycle\,24. 
\citet{2015ApJ...808L..28J} have proposed that observed weak
polar magnetic fields, and as a result the weak Cycle\,24, may have
resulted from the emergence of low-latitude flux having the opposite
polarity to that expected (which then hindered growth of the polar
fields). Predictions for Cycle\,25 are now beginning to appear 
\citep[e.g., see][]{2016SpWea..14...10P}.

Around the time of the early stages of Cycle\,24, we used helioseismic
data collected by the Birmingham Solar-Oscillations Network (BiSON) to
uncover clear signs of unusual behaviour in the near-surface layers
that appeared as far back as the latter stages of Cycle\,22 
\citep{2012ApJ...758...43B}. We analysed the cycle-induced frequency
shifts shown by modes in three different frequency bands of the p-mode
spectrum. The relationship of the frequency shifts to the ISN and the
10.7-cm radio flux -- the latter another commonly used proxy of global
solar activity -- changed noticeably, and the close correlation with
the indices was lost during a period stretching from the tail end of
Cycle\,22 through the Cycle\,23 maximum (with the lowest-frequency
modes losing the correlation first). Whilst the correlation recovered
for modes above $\simeq 2400\,\rm \mu Hz$, it failed to do so for the
lower-frequency modes. The results imply an underlying change in
structure very close to the surface, where the lower-frequency cohort
is less sensitive to perturbations. We showed that one may interpret
the structural change in terms of a thinning of the layer of
near-surface magnetic field, post-Cycle\,22.

These helioseismic markers were sufficiently robust that, with the
benefit of hindsight, they arguably may have provided an early warning
of the changes that would be seen in other proxies as the Sun headed
into Cycle\,24. At the time of writing, Cycle\,24 is about half-way
down its declining phase. Our goal in this paper is, therefore, to use
the up-to-date BiSON data to test whether the unusual behaviour
uncovered by our previous analysis has persisted through the declining
phase of the cycle, and what the results might mean for Cycle\,25. The
rest of the paper is laid out as follows. Details on the data used,
and the analysis performed, are given in Section~\ref{sec:data}. We
discuss results on the extracted frequency shifts and activity proxies
in Section~\ref{sec:res}, including a new way of presenting the
information encapsulated in the frequency shifts. We finish the paper
in Section~\ref{sec:disc} by discussing the implications of the
results, not only for Cycle\,25, but also in the wider context of the
activity behaviour of Sun-like stars.

 \section{Data and Analysis}
 \label{sec:data}

The six telescopes comprising BiSON make unresolved ``Sun-as-a-star''
observations of the visible solar disc 
\citep{1996SoPh..168....1C,2016SoPh..291....1H}, and thereby provide data that are sensitive to the globally
coherent, low angular-degree (low-$l$) solar p modes. Whilst these
modes are formed by acoustic waves that penetrate the solar core, they
are very sensitive to perturbations in the near-surface layers and
hence provide a useful diagnostic and probe of the global response of
the Sun to changing levels of solar activity and the resulting
near-surface structural changes.

The BiSON observations constitute a unique database that now stretches
over four 11-year solar activity cycles, i.e., from Cycle\,21 through
to the falling phase of the current Cycle\,24. The observations we use
here span the period 1985 July 2 through 2016 December 31, or in other words Cycles 22, 23, and most of 24; the data from Cycle 21 are too sparse to be suitable for the current analysis. The raw Doppler
velocity data were prepared for analysis using the procedures
described by 
\citet{2014MNRAS.439.2025D,2014MNRAS.441.3009D},
and then they were divided into
overlapping subsets of length 365\,days, offset by 91.25\,days. The
frequency-power spectrum of each subset was fitted to a
multi-parameter model to extract estimates of the low-$l$ mode
frequencies. 
Details can be found in
\citet{2015MNRAS.454.4120H}, but while that paper used two different codes to extract the frequencies, for this work we used only frequencies estimated using the more sophisticated method, a Markov-Chain Monte-Carlo sampler.    
 We deliberately re-analyzed the entire database in
order to verify, using independent analysis codes, the results
presented by \citet{2012ApJ...758...43B}.

Having extracted the individual frequencies, we then followed the
procedure outlined by \citet{2012ApJ...758...43B} 
 to extract averaged frequency
shifts for each 365-day segment, in three frequency bands. The bands
cover low ($1860 < \nu_{nl} \le 2400\,\rm \mu Hz$), medium ($2400 <
\nu_{nl} \le 2920\,\rm \mu Hz$) and high frequencies ($2920 < \nu_{nl}
\le 3450\,\rm \mu Hz$). The reference frequencies used for computing
the variations came from averages of the frequencies for the
{13} subsets straddling the Cycle\,22 maximum (spanning the
dates {1988 October 1 to 1992 April 30}). 



In addition to using averaged frequency shifts, we also used results
from parametrizing the frequency shifts as a function of frequency  
(\citealt{1990LNP...367..283G}; 
see also \citealt{2017MNRAS.464.4777H}).
 The frequency shifts of individual modes,
$\delta\nu_{nl}(t)$, were first scaled by the mode inertia, $E_{nl}$,
of model ``S'' of 
\cite{1996Sci...272.1286C}.
This removes
the dependence of the shifts on inertia, leaving signatures due to
perturbations in the near-surface layers. 
The scaled shifts of each
365-day subset were then fitted to a parametrized one or two-term function in
frequency $\nu$, i.e.,
 \begin{equation}
 \mathcal{F}(\nu) = a_{\rm inv} \left( \frac{\nu}{\nu_{\rm ac}} \right)^{-1} +
                    a_{\rm cub} \left( \frac{\nu}{\nu_{\rm ac}} \right)^{3},
 \end{equation}
or in the one-term case simply  
 \begin{equation}
 \mathcal{F}(\nu) =  a_{\rm cub} \left( \frac{\nu}{\nu_{\rm ac}} \right)^{3},
 \end{equation}
where $\nu_{ac}$ is the acoustic cut-off frequency of 5\,mHz. The fit
yields best-fitting coefficients  $a_{\rm cub}$, and optionally $a_{\rm inv}$,
for each subset. The rationale given by \cite{1990LNP...367..283G} for this choice of terms is, briefly, that the cubic term corresponds to a  modification of the propagation speed of the waves by a fibril magnetic field close to the surface, as also suggested by \cite{1990Natur.345..779L}, while the inverse term would correspond to a change in scale height in the superadiabatic boundary layer.

For the low-degree data we found that the $a_{\rm inv}$ term was not statistically significant, so here we work with the cubic term only.

The averaged frequency shifts for each band, and the best-fitting
coefficients from the cubic-function parametrization of the shifts,
constitute our core helioseismic diagnostic data.

 \section{Results}
 \label{sec:res}


\begin{figure*}
\includegraphics[width=0.8\linewidth]{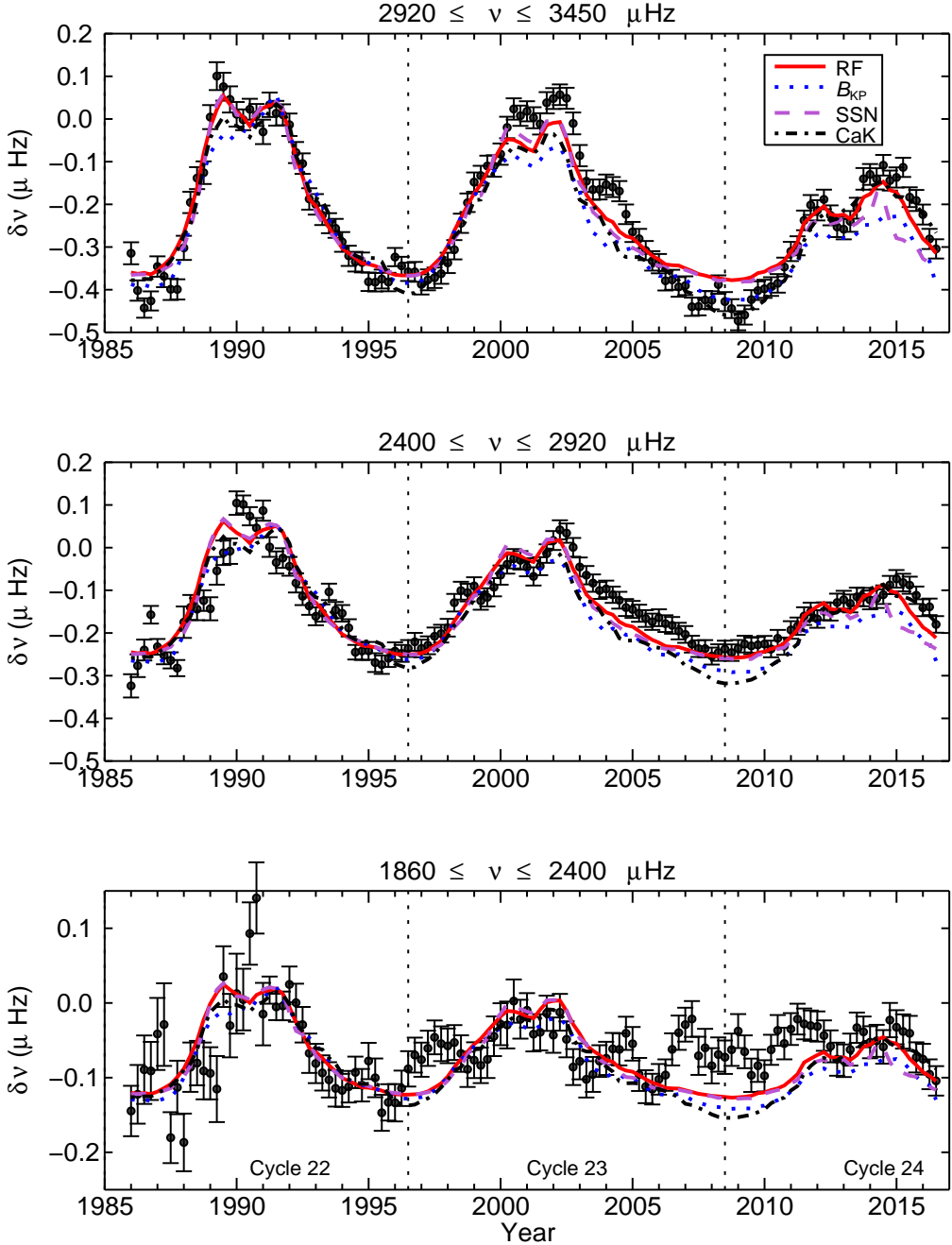}

 \caption{Averaged frequency shifts (symbols with error bars) in each of three frequency bands as a function of
   time. Also plotted are data on the 10.7-cm radio
   flux (solid red curve), CaK index (black dash-dotted curve), Kitt Peak global magnetic field strength index (dotted blue curve) and ISN (purple dashed curve), all scaled to the frequency shifts from the maximum and descending phase of Cycle 22 (see text). The vertical dotted lines indicate the cycle minima.}
 \label{fig:fig1}
\end{figure*}



\begin{figure*}

 \includegraphics[width=0.8\linewidth]{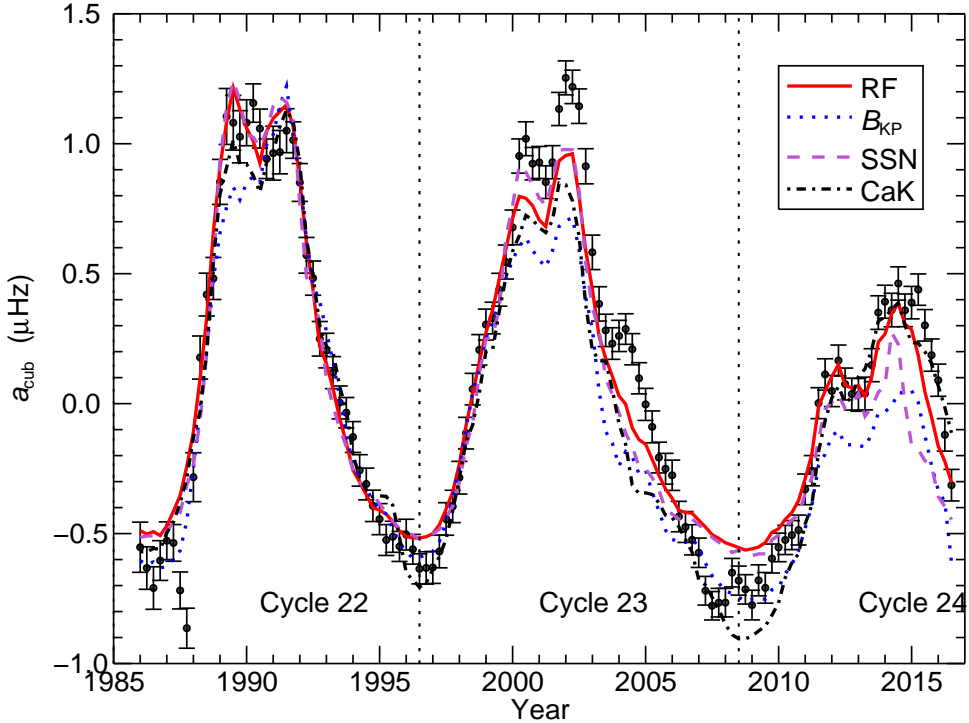}

 \caption{Extracted coefficients $a_{\mathrm{cubic}}$, from
   the cubic parametrization of the frequency shifts, as a function
   of time. Also plotted are data on the 10.7-cm radio
   flux (red solid curve), CaK index (black dash-dotted curve), global magnetic field strength index (blue dotted curve) and ISN (purple dashed curve), scaled to the frequency shifts in Cycle 22 (see text). The vertical dotted lines indicate the cycle minima.}

 \label{fig:fig2}
\end{figure*}


Fig.~\ref{fig:fig1} shows the extracted averaged frequency shifts in
each band as a function of time. Also plotted are data on the 10.7-cm
radio flux \citep{2013SpWea..11..394T}
\footnote{available from the National Geophysical Data
Center, http://www.ngdc.noaa.gov}, the revised Brussels-Locarno Sunspot Number \citep{Clette2016}\footnote{available from  http://www.sidc.be}, a merged CaK index\citep{Bertello2016}\footnote{http://solis.nso.edu/0/iss/sp\_iss.dat}   and a global magnetic field-strength index based on Kitt Peak synoptic magnetogram data \citep[see][for details]{2017MNRAS.464.4777H}, each averaged over the same
epochs as the frequencies, and scaled by a linear fit to the frequency shifts over 
the maximum and descending phase of Cycle\,22, (i.e, the date range from 1990.0 to 1996.5). Fig.~\ref{fig:fig2} shows the extracted coefficients
$a_{\rm cub}$, as a function of time, with the
radio-flux, sunspot, CaK, and magnetic data overplotted, again scaled by a linear fit to the cubic coefficients in the latter part of Cycle 22.

\begin{figure*}
 \includegraphics[width=0.8\linewidth]{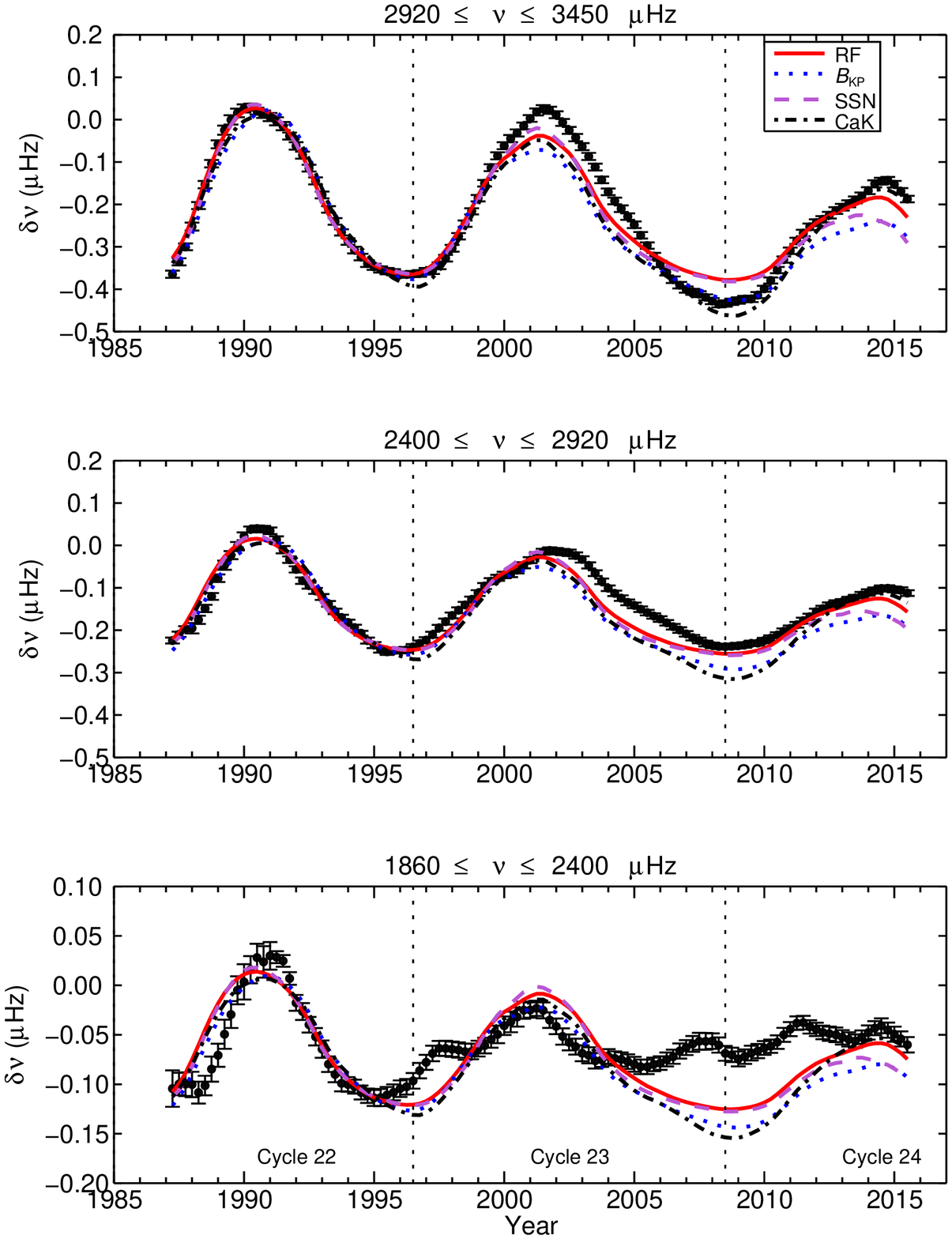}
 \caption{Averaged frequency shifts (symbols) in each of three frequency bands as a function of
   time, but after smoothing to remove shorter-term variations. Also plotted are data on the 10.7-cm
   radio flux (red solid curve),  ISN (purple dashed curve), CaK index (black dash-dotted curve)  and global magnetic index (blue dotted curve), all scaled by a linear fit to the smoothed frequency shifts over the maximum and descending phase of Solar Cycle 22 (see text). The vertical dotted lines indicate the cycle minima.}

 \label{fig:fig3}
\end{figure*}

\begin{figure*}

 \includegraphics[width=0.8\linewidth]{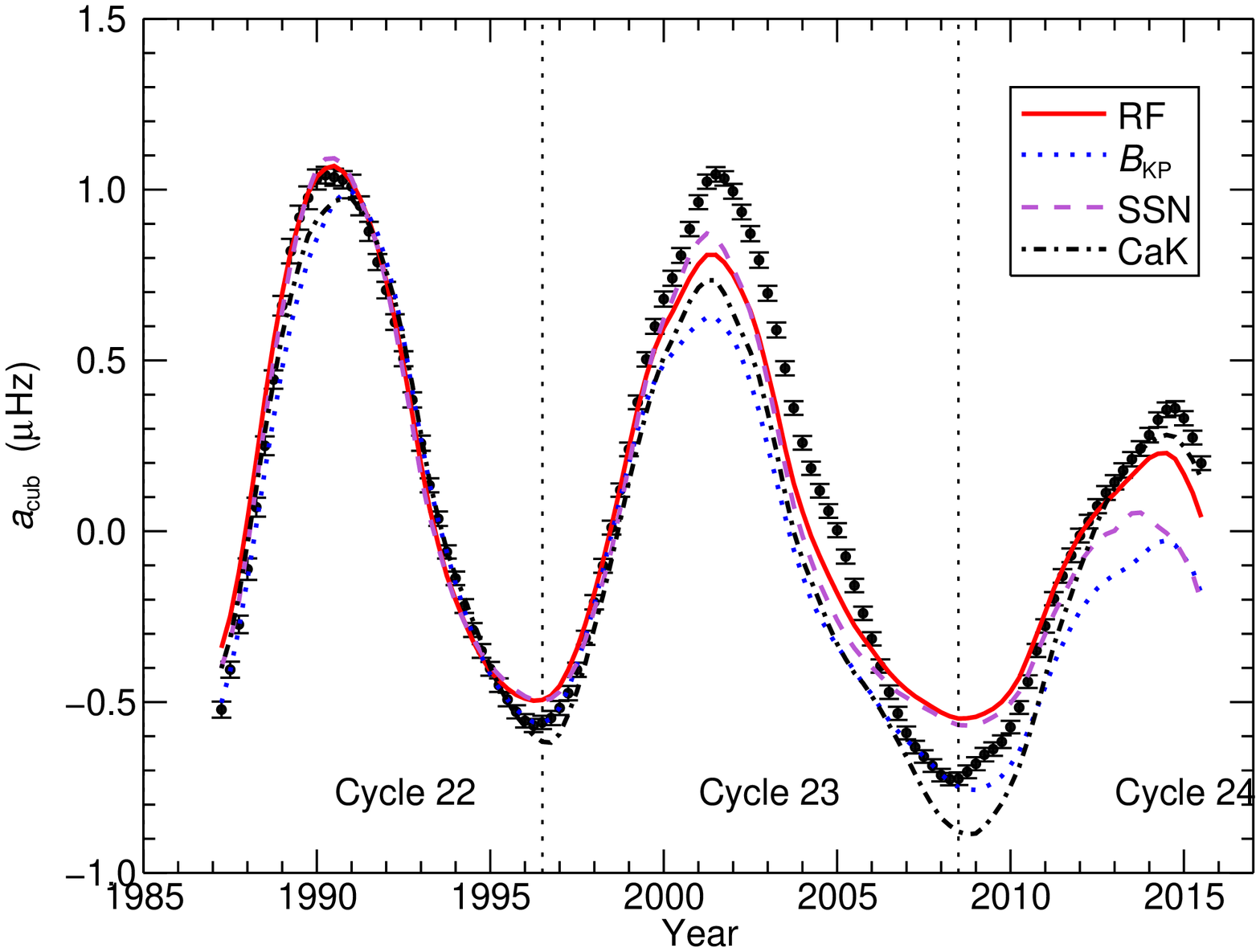}

 \caption{Coefficient $a_{\rm cubic}$, as a function
   of time but after smoothing the data to remove shorter-term
   variations. Also plotted are data on the 10.7-cm
   radio flux (red solid curve),  CaK index (black dash-dotted curve), ISN (purple dashed line), and magnetic global index (blue dotted curve) (see text), scaled by a linear fit to the smoothed  $a_{\rm cubic}$ coefficients over the maximum and descending phase of Solar Cycle 22. The vertical dotted lines indicate the cycle minima.}

 \label{fig:fig4}
\end{figure*}

Signatures of the 11-year solar activity cycle are the dominant
feature of both sets of diagnostics. However, both frequencies and activity proxies also show some
shorter-term variability with an average period of around
2\,years. This periodicity is a known feature of several global
proxies of solar activity 
\citep[e.g., see][]{2014SSRv..186..359B}
and
has been detected in a number of helioseismic datasets 
\citep[e.g., see][]{2010ApJ...718L..19F,2012MNRAS.420.1405B,2012A&A...539A.135S,2013ApJ...765..100S}.
Following 
\citet{2012ApJ...758...43B}, we have removed shorter-period
variations, including the 2-year signal, by smoothing the respective
sets of averaged frequency shifts and extracted coefficients over nine samples or 2.25 years.  The
resulting smoothed diagnostic data are plotted in Fig.~\ref{fig:fig3} for the average frequency shifts and Fig.~\ref{fig:fig4} for the cubic coefficients, together with smoothed and scaled versions of the activity proxies.


Figures~\ref{fig:fig1} and ~\ref{fig:fig3} correspond to Figures 2 and 3 of 
\citet{2012ApJ...758...43B}, with the slight difference that we have scaled the activity proxies to the frequencies over a wider range in time, while the cubic-fit coefficients in Figures~\ref{fig:fig2} and \ref{fig:fig4} combine data from all three bands but are more heavily weighted towards the higher-frequency modes.
The first important point to make is that our re-analysis of the BiSON
data confirms the results presented in 
\citet{2012ApJ...758...43B},
although there are some differences in detail that may be due to the improved frequency 
estimation. Prior to
$\simeq 1994$, during the latter stages of Cycle\,22, changes in the
frequencies followed reasonably closely the variations shown by the
global activity proxies. However, at epochs thereafter, the behaviour
of the frequency shifts in the low-frequency band departed strongly
from the proxies, with detected variations in the frequencies being
much weaker than expected (based on the behaviour seen in Cycles\,21
and 22); in particular, these frequencies stayed higher than expected during the very low-activity period of the minimum following Cycle 23.  We also note that the relationship between the radio flux
and sunspot data temporarily changed during the declining phase of Cycle\,23 (as pointed out by
\citealt{2011SoPh..272..337T}; see also \citealt{Clette2016} for comments on the comparison with the newly recalibrated sunspot data).

Our new results extend the observations into the declining phase of
Cycle\,24, and they show clearly that the significant departures seen at
low frequencies have persisted as we head towards the onset of the
next cycle. Put another way, the acoustic properties of the
near-surface layers have failed to re-set to their pre-1994 state. 
While for the mid- and high-frequency band shifts and the cubic parametrization coefficient the correlation with the activity indices does appear to recover in the rising phase of Cycle 24, the shifts again deviate from the extrapolated fit to Cycle 22 in the most recent data corresponding to the declining phase of Cycle\,24, at least for the RF, magnetic, and sunspot indices. We note that the scaled RF and sunspot proxies show very similar behaviour except in the declining phase of Cycle 24, while the CaK and magnetic proxies are fairly close to one another. While the magnetic and CaK proxies fall to noticeably lower levels during the Cycle 23/24 minimum than during the previous minimum, the difference between the minima is less pronounced for the RF and sunspot number. Interestingly, the frequency shifts in the high-frequency band, as well as the cubic-fit coefficients, appear to follow the magnetic and CaK pattern, while the frequency shifts in the middle band follow the RF and sunspot number and the low-frequency band is not a good match to any of the proxies in this period. During the maximum epoch of Cycle 24 the frequencies in the high and middle bands seem to follow the extrapscaled RF and CaK proxies while the sunspot and magnetic proxies show poorer agreement, and in the declining phase so far only the CaK looks like a good match to the frequency shifts. We emphasise that all of the proxies have been scaled to match the relationship to the frequency shifts that was seen in the 1990\,--\,1996.5 epoch; better fits could be obtained by fitting to each cycle separately. The difference in behaviour at the Cycle 23/4 minimum is striking, however, and it is not an artefact of the scaling. \citet{2017arXiv170203149B} also found that the medium-degree frequencies from the Global Oscillation Network Group were systematically lower during this minimum than during the previous one, as would be expected when the fields are weaker.

\begin{figure*}

\includegraphics[width=\linewidth]{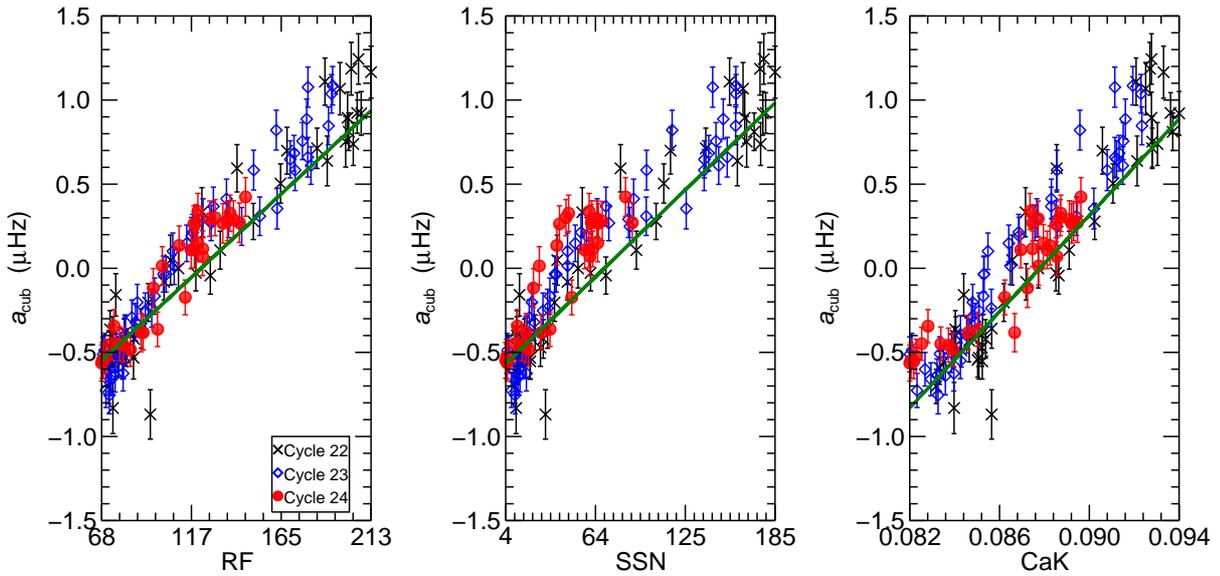}
\caption{Cubic coefficient for $l=0$ only, plotted against different global activity indices, for solar cycles 22 (black crosses), 23 (blue open diamonds), and 24 (red filled circles). The solid line indicates a linear fit to the 
data between 1990 and 1996, i.e. the maximum and descending phase of Cycle 22.}
\label{fig:rhfig5}
\end{figure*}

Because the BiSON observations are primarily sensitive to the sectoral, or $m=l$ modes, we can expect the frequency shifts for $l\geq 1$ to show some hysteresis with global activity measures. To examine the impact this may have on our results, we carried out the frequency-cubed parametrization for each $l$ separately. Figure~\ref{fig:rhfig5} shows the unsmoothed cubic coefficient derived from fits to the  $l=0$ shifts only, plotted against each of the three unresolved activity indices and colour-coded by cycle. Also shown is a straight line representing the extrapolation of a linear fit to the Cycle 22 data after 1990. In each case, it can be seen that the data for the Cycle 23 and 24 maxima lie generally above the line. This suggests that in the two most recent cycles we are seeing a {\em larger} shift overall for the same amount of activity than we did in Cycle 22, in the middle and higher frequency bands that dominate the cubic fit. This might make sense if there has been a shift towards a greater proportion of weak, ephemeral activity that does not register in the sunspot number or on the synoptic magnetic charts but could still have an influence on the modes. The numerical results of fits between  the $l=0$ cubic coefficient and the activity proxies, for each cycle individually and for the entire data sets, are given in Table~\ref{tab:l0fit}.

\begin{figure*}
\includegraphics[width=0.8\linewidth]{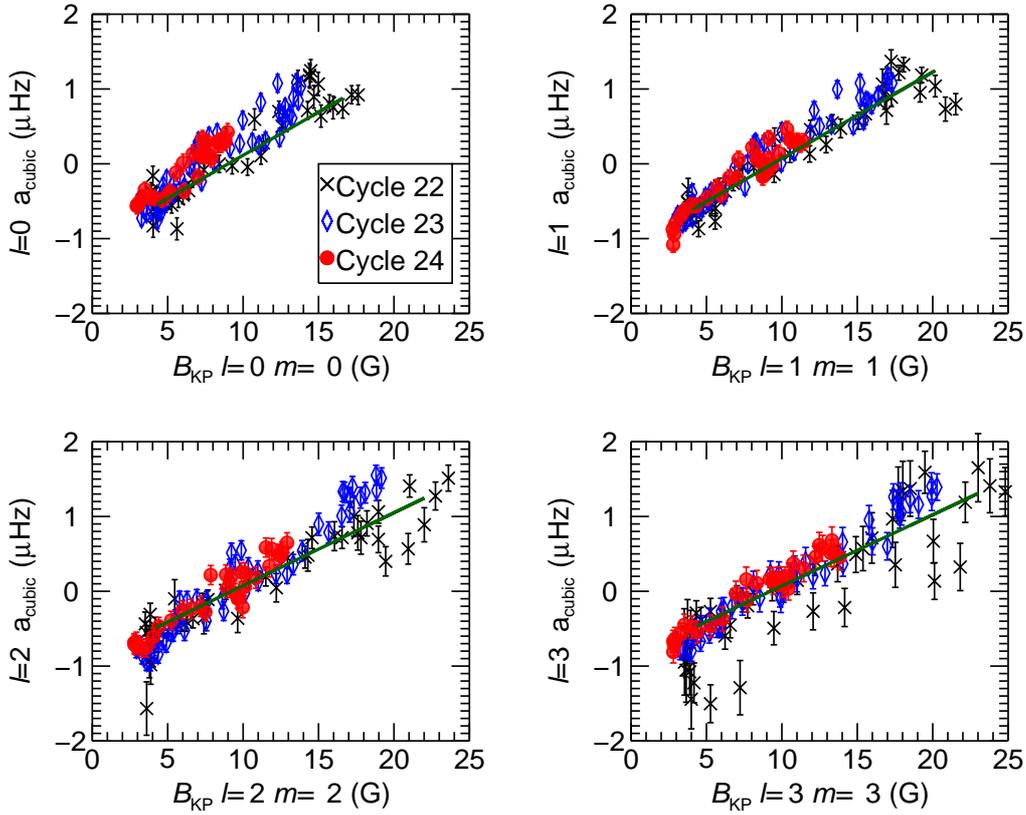}
\caption{Cubic coefficients for fits to MCMC scaled frequency shifts for individual $l$ plotted as a function of the $l,m=l$ projection of the 
magnetic field strength, for solar cycles 22 (black crosses), 23 (blue open diamonds), and 24 (red filled circles).}
\label{fig:rhfig6} 
\end{figure*}

In the case of the magnetic proxy, we do have information on the latitudinal distribution of activity in the synoptic magnetograms, and we can use this to derive an index appropriate to the sectoral mode at each value of $l$ \citep[see, for example,][]{2004MNRAS.352.1102C}. In Figure~\ref{fig:rhfig6} we 
show the cubic coefficients  for each $l$ separately as a function of the projection of the latitudinal field-strength distribution on the corresponding
$m=l$ spherical harmonic. The results of linear fits to the magnetic proxy for each cycle separately and for the whole dataset are shown in Table~\ref{tab:magfit}. Again, we can see that the points corresponding to the maxima of Cycles 23 and 24 lie above the curve indicating the trend in Cycle 22, so we cannot attribute the hysteresis we observe in the averaged frequencies simply to the latitudinal distribution of activity. The $l=3$ points from early Cycle 22 that lie well below the line are probably due to poor fits of these modes (to which a Sun-as-a-star instrument like BiSON is not very sensitive) in the low duty cycle of the early observations.

\begin{table*}
\caption{Results of linear fits between global proxies and cubic frequency-shift coefficient for $l=0$, for the three cycles individually and the whole period. These fits were carried out for non-overlapping data only.}
\label{tab:l0fit}
\begin{tabular}{llrrrrr}
\hline
Proxy & Period &  Intercept ($\mu$Hz) & Slope ($\mu$Hz/activity unit) & $\chi^2$ & $R$ \\
\hline
RF & Cycle 22 & $( -1.266\pm 0.096)\times 10^{0}$ & 
$( 1.080\pm 0.073)\times 10^{-2}$ &       2.59160  &      0.946495\\
" & Cycle 23 & $( -1.498\pm 0.096)\times 10^{0}$ & 
$( 1.310\pm 0.077)\times 10^{-2}$ &       1.13799 &      0.980589\\
" &  Cycle 24 &$( -1.323\pm 0.163)\times 10^{0}$ & 
$( 1.197\pm 0.151)\times 10^{-2}$ &      0.725496 &     0.967366\\
" &  All &$( -1.344\pm 0.060)\times 10^{0}$ & $( 1.178\pm 0.049)\times 10^{-2}$
 &       1.59291 &     0.958641\\
\hline
SSN & Cycle 22 & $( -5.675\pm 0.560)\times 10^{-1}$ & 
$( 8.866\pm 0.601)\times 10^{-3}$ &       2.82573  &      0.941283\\
" & Cycle 23 & $( -5.982\pm 0.496)\times 10^{-1}$ & 
$( 1.003\pm 0.060)\times 10^{-2}$ &       2.14756 &      0.963345\\
" &  Cycle 24 &$( -4.994\pm 0.718)\times 10^{-1}$ & 
$( 1.137\pm 0.161)\times 10^{-2}$ &       2.88601 &     0.858331\\
" &  All &$( -5.227\pm 0.308)\times 10^{-1}$ & $( 9.204\pm 0.397)\times 10^{-3}$
 &       2.75214 &     0.933063\\
\hline
CaK & Cycle 22 & $( -1.368\pm 0.093)\times 10^{1}$ & 
$( 1.563\pm 0.106)\times 10^{2}$ &       2.87813  &      0.943093\\
" & Cycle 23 & $( -1.497\pm 0.089)\times 10^{1}$ & 
$( 1.728\pm 0.102)\times 10^{2}$ &       1.35981 &      0.975541\\
" &  Cycle 24 &$( -1.030\pm 0.129)\times 10^{1}$ & 
$( 1.185\pm 0.150)\times 10^{2}$ &      0.780809 &     0.965822\\
" &  All &$( -1.327\pm 0.056)\times 10^{1}$ & $( 1.526\pm 0.065)\times 10^{2}$
 &       2.14424 &     0.946690\\
\hline

\end{tabular}
\end{table*}

\begin{table*}
\caption{Results of linear fits between $m=l$ components of the Kitt Peak magnetic index and the cubic frequency-shift coefficient for each l, for the three cycles individually and the whole period.}
\label{tab:magfit}
\begin{tabular}{llrrrrr}
\hline
$l$ & Period &  Intercept ($\mu$Hz) & Slope ($\mu$Hz/G) & $\chi^2$ & $R$ \\
\hline
0 & Cycle 22 & $ -0.990\pm 0.081$ & $ 0.119\pm 0.081$ & 4.153 & 0.918 \\
0 & Cycle 23 & $ -1.111\pm 0.076$ & $ 0.147\pm 0.076$ & 2.596 & 0.954 \\
0 & Cycle 24 & $ -0.944\pm 0.117$ & $ 0.143\pm 0.117$ & 0.842 & 0.963 \\
0 & All  & $ -0.950\pm 0.047$ & $ 0.126\pm 0.006$ & 3.187 & 0.921\\
1 & Cycle 22 & $ -1.025\pm 0.072$ & $ 0.112\pm 0.072$ & 2.588 & 0.959 \\
1 & Cycle 23 & $ -1.044\pm 0.067$ & $ 0.124\pm 0.067$ & 2.614 & 0.963 \\
1 & Cycle 24 & $ -1.237\pm 0.096$ & $ 0.144\pm 0.096$ & 2.858 & 0.940 \\
1 & All  & $ -1.063\pm 0.042$ & $ 0.121\pm 0.004$ & 2.627 & 0.955\\
2 & Cycle 22 & $ -0.953\pm 0.090$ & $ 0.101\pm 0.090$ & 0.855 & 0.965 \\
2 & Cycle 23 & $ -1.299\pm 0.080$ & $ 0.145\pm 0.080$ & 3.575 & 0.958 \\
2 & Cycle 24 & $ -1.084\pm 0.113$ & $ 0.127\pm 0.113$ & 2.056 & 0.938 \\
2 & All  & $ -1.106\pm 0.051$ & $ 0.124\pm 0.005$ & 2.629 & 0.938\\
3 & Cycle 22 & $ -1.085\pm 0.149$ & $ 0.107\pm 0.149$ & 2.430 & 0.913 \\
3 & Cycle 23 & $ -1.118\pm 0.100$ & $ 0.122\pm 0.100$ & 1.079 & 0.969 \\
3 & Cycle 24 & $ -1.047\pm 0.133$ & $ 0.122\pm 0.133$ & 0.836 & 0.966 \\
3 & All  & $ -1.072\pm 0.068$ & $ 0.117\pm 0.007$ & 1.433 & 0.917\\
\hline

\end{tabular}
\end{table*}

 \section{Discussion}
 \label{sec:disc}


We have used the latest BiSON helioseismic data to show that
previously uncovered changes in the structure of the near-surface
layers of the Sun, which date back to the latter stages of Cycle\,22
(around 1994), have persisted through the declining phase of the
current, weak Cycle\,24. The acoustic properties have as such failed
to re-set to their pre-1994 state. While the agreement at higher frequencies did
appear to recover in the rising phase of Cycle 24, when the most recent 
data are added we can see that there are again differences from the proxies extrapolated from Cycle 22. This supports the suggestion of \citet{2012ApJ...758...43B} that the magnetic changes affecting the Cycle 23 (and later) oscillation frequencies were confined to a thinner layer than those in Cycle 22.


We also find that the sensitivity of the frequencies in the higher-frequency bands to the magnetic proxy is slightly higher in the two most recent cycles than in Cycle\,22, which could be due to a higher proportion of weaker, more ephemeral active regions that are not accounted for in the synoptic magnetic data. This observation still holds when we separate out the frequency shifts by degree and compare with a latitudinally resolved magnetic proxy. 

It is tempting to speculate whether these results, and the multitude
of other unusual signatures relating to Cycle\,24, might be indicative
of a longer-lasting transition in solar activity behaviour, and the
operation of the solar dynamo. 

The existence of the Maunder Minimum, and other similar minima
suggested by proxy data relevant to millennal timescales, indicate
that there have likely been periods when the action of the dynamo has
been altered significantly. We finish by speculating whether these
events might presage a radical transition suggested by data on other
stars. Results on activity cycle periods shown by other stars hint at
a change in cycle behaviour -- a possible transition from one type of
dynamo action to another -- at a surface rotation period of around
20\,days \citep{2007ApJ...657..486B}. 
There is also more recent intriguing
evidence from asteroseismic results on solar-type stars 
\citep{2016Natur.529..181V,2016ApJ...826L...2M}
that shows that the spin-down behaviour of cool stars
changes markedly once they reach a critical epoch, with the
corresponding surface rotation period depending on stellar mass. For
solar-mass stars, the results suggest a change in behaviour at about
the solar age (and solar rotation period).

\subsection*{ACKNOWLEDGMENTS}

We would like to thank al those who are, or have been, associated
with BiSON, in particular P.~Pall\'e and T.~Roca-Cortes in Tenerife
and E.~Rhodes~Jr. and S.~Pinkerton at Mt.~Wilson.  BiSON is funded by
the Science and Technology Facilities Council (STFC), under grant ST/M00077X/1. 
SJH, GRD, YPE, and RH acknowledge the support of the UK Science and
Technology Facilities Council (STFC). Funding for the Stellar Astrophysics
Centre (SAC) is provided by The Danish National Research Foundation
(Grant DNRF106). 
NSO/Kitt Peak data used here were produced cooperatively by
NSF/NOAO, NASA/GSFC, and NOAA/SEL; SOLIS data are
produced cooperatively by NSF/NSO and NASA/LWS.
RH thanks the National Solar Observatory for computing support. SB acknowledges National Science Foundation (NSF) grant AST-1514676.

\bibliography{mntrans}

\begin{thebibliography}{}
\makeatletter
\relax
\def\mn@urlcharsother{\let\do\@makeother \do\$\do\&\do\#\do\^\do\_\do\%\do\~}
\def\mn@doi{\begingroup\mn@urlcharsother \@ifnextchar [ {\mn@doi@}
  {\mn@doi@[]}}
\def\mn@doi@[#1]#2{\def\@tempa{#1}\ifx\@tempa\@empty \href
  {http://dx.doi.org/#2} {doi:#2}\else \href {http://dx.doi.org/#2} {#1}\fi
  \endgroup}
\def\mn@eprint#1#2{\mn@eprint@#1:#2::\@nil}
\def\mn@eprint@arXiv#1{\href {http://arxiv.org/abs/#1} {{\tt arXiv:#1}}}
\def\mn@eprint@dblp#1{\href {http://dblp.uni-trier.de/rec/bibtex/#1.xml}
  {dblp:#1}}
\def\mn@eprint@#1:#2:#3:#4\@nil{\def\@tempa {#1}\def\@tempb {#2}\def\@tempc
  {#3}\ifx \@tempc \@empty \let \@tempc \@tempb \let \@tempb \@tempa \fi \ifx
  \@tempb \@empty \def\@tempb {arXiv}\fi \@ifundefined
  {mn@eprint@\@tempb}{\@tempb:\@tempc}{\expandafter \expandafter \csname
  mn@eprint@\@tempb\endcsname \expandafter{\@tempc}}}

\bibitem[\protect\citeauthoryear{{Basu}, {Broomhall}, {Chaplin}  \&
  {Elsworth}}{{Basu} et~al.}{2012}]{2012ApJ...758...43B}
{Basu} S.,  {Broomhall} A.-M.,  {Chaplin} W.~J.,   {Elsworth} Y.,  2012,
  \mn@doi [\apj] {10.1088/0004-637X/758/1/43}, \href
  {http://adsabs.harvard.edu/abs/2012ApJ...758...43B} {758, 43}

\bibitem[\protect\citeauthoryear{{Bazilevskaya}, {Broomhall}, {Elsworth}  \&
  {Nakariakov}}{{Bazilevskaya} et~al.}{2014}]{2014SSRv..186..359B}
{Bazilevskaya} G.,  {Broomhall} A.-M.,  {Elsworth} Y.,   {Nakariakov} V.~M.,
  2014, \mn@doi [\ssr] {10.1007/s11214-014-0068-0}, \href
  {http://adsabs.harvard.edu/abs/2014SSRv..186..359B} {186, 359}

\bibitem[\protect\citeauthoryear{Bertello, Pevtsov, Tlatov  \& Singh}{Bertello
  et~al.}{2016}]{Bertello2016}
Bertello L.,  Pevtsov A.,  Tlatov A.,   Singh J.,  2016, \mn@doi [Solar
  Physics] {10.1007/s11207-016-0927-9}, 291, 2967

\bibitem[\protect\citeauthoryear{{B{\"o}hm-Vitense}}{{B{\"o}hm-Vitense}}{2007}]{2007ApJ...657..486B}
{B{\"o}hm-Vitense} E.,  2007, \mn@doi [\apj] {10.1086/510482}, \href
  {http://adsabs.harvard.edu/abs/2007ApJ...657..486B} {657, 486}

\bibitem[\protect\citeauthoryear{{Broomhall}}{{Broomhall}}{2017}]{2017arXiv170203149B}
{Broomhall} A.-M.,  2017, preprint, \href
  {http://adsabs.harvard.edu/abs/2017arXiv170203149B} {} (\mn@eprint {arXiv}
  {1702.03149})

\bibitem[\protect\citeauthoryear{{Broomhall}, {Chaplin}, {Elsworth}  \&
  {Simoniello}}{{Broomhall} et~al.}{2012}]{2012MNRAS.420.1405B}
{Broomhall} A.-M.,  {Chaplin} W.~J.,  {Elsworth} Y.,   {Simoniello} R.,  2012,
  \mn@doi [\mnras] {10.1111/j.1365-2966.2011.20123.x}, \href
  {http://adsabs.harvard.edu/abs/2012MNRAS.420.1405B} {420, 1405}

\bibitem[\protect\citeauthoryear{{Chaplin} et~al.,}{{Chaplin}
  et~al.}{1996}]{1996SoPh..168....1C}
{Chaplin} W.~J.,  et~al., 1996, \mn@doi [\solphys] {10.1007/BF00145821}, \href
  {http://adsabs.harvard.edu/abs/1996SoPh..168....1C} {168, 1}

\bibitem[\protect\citeauthoryear{{Chaplin}, {Elsworth}, {Isaak}, {Miller}  \&
  {New}}{{Chaplin} et~al.}{2004}]{2004MNRAS.352.1102C}
{Chaplin} W.~J.,  {Elsworth} Y.,  {Isaak} G.~R.,  {Miller} B.~A.,   {New} R.,
  2004, \mn@doi [\mnras] {10.1111/j.1365-2966.2004.07998.x}, \href
  {http://adsabs.harvard.edu/abs/2004MNRAS.352.1102C} {352, 1102}

\bibitem[\protect\citeauthoryear{{Christensen-Dalsgaard}
  et~al.,}{{Christensen-Dalsgaard} et~al.}{1996}]{1996Sci...272.1286C}
{Christensen-Dalsgaard} J.,  et~al., 1996, \mn@doi [Science]
  {10.1126/science.272.5266.1286}, \href
  {http://adsabs.harvard.edu/abs/1996Sci...272.1286C} {272, 1286}

\bibitem[\protect\citeauthoryear{{Clette} \& {Lef{\`e}vre}}{{Clette} \&
  {Lef{\`e}vre}}{2012}]{2012JSWSC...2A..06C}
{Clette} F.,  {Lef{\`e}vre} L.,  2012, \mn@doi [Journal of Space Weather and
  Space Climate] {10.1051/swsc/2012007}, \href
  {http://adsabs.harvard.edu/abs/2012JSWSC...2A..06C} {2, A06}

\bibitem[\protect\citeauthoryear{Clette, Lef{\`e}vre, Cagnotti, Cortesi  \&
  Bulling}{Clette et~al.}{2016}]{Clette2016}
Clette F.,  Lef{\`e}vre L.,  Cagnotti M.,  Cortesi S.,   Bulling A.,  2016,
  \mn@doi [Solar Physics] {10.1007/s11207-016-0875-4}, 291, 2733

\bibitem[\protect\citeauthoryear{{Davies}, {Broomhall}, {Chaplin}, {Elsworth}
  \& {Hale}}{{Davies} et~al.}{2014a}]{2014MNRAS.439.2025D}
{Davies} G.~R.,  {Broomhall} A.~M.,  {Chaplin} W.~J.,  {Elsworth} Y.,   {Hale}
  S.~J.,  2014a, \mn@doi [\mnras] {10.1093/mnras/stu080}, \href
  {http://adsabs.harvard.edu/abs/2014MNRAS.439.2025D} {439, 2025}

\bibitem[\protect\citeauthoryear{{Davies}, {Chaplin}, {Elsworth}  \&
  {Hale}}{{Davies} et~al.}{2014b}]{2014MNRAS.441.3009D}
{Davies} G.~R.,  {Chaplin} W.~J.,  {Elsworth} Y.,   {Hale} S.~J.,  2014b,
  \mn@doi [\mnras] {10.1093/mnras/stu803}, \href
  {http://adsabs.harvard.edu/abs/2014MNRAS.441.3009D} {441, 3009}

\bibitem[\protect\citeauthoryear{{Fletcher}, {Broomhall}, {Salabert}, {Basu},
  {Chaplin}, {Elsworth}, {Garcia}  \& {New}}{{Fletcher}
  et~al.}{2010}]{2010ApJ...718L..19F}
{Fletcher} S.~T.,  {Broomhall} A.-M.,  {Salabert} D.,  {Basu} S.,  {Chaplin}
  W.~J.,  {Elsworth} Y.,  {Garcia} R.~A.,   {New} R.,  2010, \mn@doi [\apjl]
  {10.1088/2041-8205/718/1/L19}, \href
  {http://adsabs.harvard.edu/abs/2010ApJ...718L..19F} {718, L19}

\bibitem[\protect\citeauthoryear{{Gough}}{{Gough}}{1990}]{1990LNP...367..283G}
{Gough} D.~O.,  1990, in {Osaki} Y.,  {Shibahashi} H.,  eds,  Lecture Notes in
  Physics, Berlin Springer Verlag Vol. 367, Progress of Seismology of the Sun
  and Stars. p.~283, \mn@doi{10.1007/3-540-53091-6}

\bibitem[\protect\citeauthoryear{{Hale}, {Howe}, {Chaplin}, {Davies}  \&
  {Elsworth}}{{Hale} et~al.}{2016}]{2016SoPh..291....1H}
{Hale} S.~J.,  {Howe} R.,  {Chaplin} W.~J.,  {Davies} G.~R.,   {Elsworth}
  Y.~P.,  2016, \mn@doi [\solphys] {10.1007/s11207-015-0810-0}, \href
  {http://adsabs.harvard.edu/abs/2016SoPh..291....1H} {291, 1}

\bibitem[\protect\citeauthoryear{{Hathaway}}{{Hathaway}}{2015}]{2015LRSP...12....4H}
{Hathaway} D.~H.,  2015, \mn@doi [Living Reviews in Solar Physics]
  {10.1007/lrsp-2015-4}, \href
  {http://adsabs.harvard.edu/abs/2015LRSP...12....4H} {12, 4}

\bibitem[\protect\citeauthoryear{{Hathaway} \& {Rightmire}}{{Hathaway} \&
  {Rightmire}}{2010}]{2010Sci...327.1350H}
{Hathaway} D.~H.,  {Rightmire} L.,  2010, \mn@doi [Science]
  {10.1126/science.1181990}, \href
  {http://adsabs.harvard.edu/abs/2010Sci...327.1350H} {327, 1350}

\bibitem[\protect\citeauthoryear{{Howe}, {Christensen-Dalsgaard}, {Hill},
  {Komm}, {Larson}, {Rempel}, {Schou}  \& {Thompson}}{{Howe}
  et~al.}{2013}]{2013ApJ...767L..20H}
{Howe} R.,  {Christensen-Dalsgaard} J.,  {Hill} F.,  {Komm} R.,  {Larson}
  T.~P.,  {Rempel} M.,  {Schou} J.,   {Thompson} M.~J.,  2013, \mn@doi [\apjl]
  {10.1088/2041-8205/767/1/L20}, \href
  {http://adsabs.harvard.edu/abs/2013ApJ...767L..20H} {767, L20}

\bibitem[\protect\citeauthoryear{{Howe}, {Davies}, {Chaplin}, {Elsworth}  \&
  {Hale}}{{Howe} et~al.}{2015}]{2015MNRAS.454.4120H}
{Howe} R.,  {Davies} G.~R.,  {Chaplin} W.~J.,  {Elsworth} Y.~P.,   {Hale}
  S.~J.,  2015, \mn@doi [\mnras] {10.1093/mnras/stv2210}, \href
  {http://adsabs.harvard.edu/abs/2015MNRAS.454.4120H} {454, 4120}

\bibitem[\protect\citeauthoryear{{Howe}, {Basu}, {Davies}, {Ball}, {Chaplin},
  {Elsworth}  \& {Komm}}{{Howe} et~al.}{2017}]{2017MNRAS.464.4777H}
{Howe} R.,  {Basu} S.,  {Davies} G.~R.,  {Ball} W.~H.,  {Chaplin} W.~J.,
  {Elsworth} Y.,   {Komm} R.,  2017, \mn@doi [\mnras] {10.1093/mnras/stw2668},
  \href {http://adsabs.harvard.edu/abs/2017MNRAS.464.4777H} {464, 4777}

\bibitem[\protect\citeauthoryear{{Jiang}, {Cameron}  \&
  {Sch{\"u}ssler}}{{Jiang} et~al.}{2015}]{2015ApJ...808L..28J}
{Jiang} J.,  {Cameron} R.~H.,   {Sch{\"u}ssler} M.,  2015, \mn@doi [\apjl]
  {10.1088/2041-8205/808/1/L28}, \href
  {http://adsabs.harvard.edu/abs/2015ApJ...808L..28J} {808, L28}

\bibitem[\protect\citeauthoryear{{Libbrecht} \& {Woodard}}{{Libbrecht} \&
  {Woodard}}{1990}]{1990Natur.345..779L}
{Libbrecht} K.~G.,  {Woodard} M.~F.,  1990, \mn@doi [\nat] {10.1038/345779a0},
  \href {http://adsabs.harvard.edu/abs/1990Natur.345..779L} {345, 779}

\bibitem[\protect\citeauthoryear{{Livingston}, {Penn}  \&
  {Svalgaard}}{{Livingston} et~al.}{2012}]{2012ApJ...757L...8L}
{Livingston} W.,  {Penn} M.~J.,   {Svalgaard} L.,  2012, \mn@doi [\apjl]
  {10.1088/2041-8205/757/1/L8}, \href
  {http://adsabs.harvard.edu/abs/2012ApJ...757L...8L} {757, L8}

\bibitem[\protect\citeauthoryear{{Metcalfe}, {Egeland}  \& {van
  Saders}}{{Metcalfe} et~al.}{2016}]{2016ApJ...826L...2M}
{Metcalfe} T.~S.,  {Egeland} R.,   {van Saders} J.,  2016, \mn@doi [\apjl]
  {10.3847/2041-8205/826/1/L2}, \href
  {http://adsabs.harvard.edu/abs/2016ApJ...826L...2M} {826, L2}

\bibitem[\protect\citeauthoryear{{Pesnell}}{{Pesnell}}{2008}]{2008SoPh..252..209P}
{Pesnell} W.~D.,  2008, \mn@doi [\solphys] {10.1007/s11207-008-9252-2}, \href
  {http://adsabs.harvard.edu/abs/2008SoPh..252..209P} {252, 209}

\bibitem[\protect\citeauthoryear{{Pesnell}}{{Pesnell}}{2012}]{2012SoPh..281..507P}
{Pesnell} W.~D.,  2012, \mn@doi [\solphys] {10.1007/s11207-012-9997-5}, \href
  {http://adsabs.harvard.edu/abs/2012SoPh..281..507P} {281, 507}

\bibitem[\protect\citeauthoryear{{Pesnell}}{{Pesnell}}{2016}]{2016SpWea..14...10P}
{Pesnell} W.~D.,  2016, \mn@doi [Space Weather] {10.1002/2015SW001304}, \href
  {http://adsabs.harvard.edu/abs/2016SpWea..14...10P} {14, 10}

\bibitem[\protect\citeauthoryear{{Salabert}, {Garc{\'{\i}}a}  \&
  {Turck-Chi{\`e}ze}}{{Salabert} et~al.}{2015}]{2015A&A...578A.137S}
{Salabert} D.,  {Garc{\'{\i}}a} R.~A.,   {Turck-Chi{\`e}ze} S.,  2015, \mn@doi
  [\aap] {10.1051/0004-6361/201425236}, \href
  {http://adsabs.harvard.edu/abs/2015A%26A...578A.137S} {578, A137}

\bibitem[\protect\citeauthoryear{{Simoniello}, {Finsterle}, {Salabert},
  {Garc{\'{\i}}a}, {Turck-Chi{\`e}ze}, {Jim{\'e}nez}  \& {Roth}}{{Simoniello}
  et~al.}{2012}]{2012A&A...539A.135S}
{Simoniello} R.,  {Finsterle} W.,  {Salabert} D.,  {Garc{\'{\i}}a} R.~A.,
  {Turck-Chi{\`e}ze} S.,  {Jim{\'e}nez} A.,   {Roth} M.,  2012, \mn@doi [\aap]
  {10.1051/0004-6361/201118057}, \href
  {http://adsabs.harvard.edu/abs/2012A%26A...539A.135S} {539, A135}

\bibitem[\protect\citeauthoryear{{Simoniello}, {Jain}, {Tripathy},
  {Turck-Chi{\`e}ze}, {Baldner}, {Finsterle}, {Hill}  \& {Roth}}{{Simoniello}
  et~al.}{2013}]{2013ApJ...765..100S}
{Simoniello} R.,  {Jain} K.,  {Tripathy} S.~C.,  {Turck-Chi{\`e}ze} S.,
  {Baldner} C.,  {Finsterle} W.,  {Hill} F.,   {Roth} M.,  2013, \mn@doi [\apj]
  {10.1088/0004-637X/765/2/100}, \href
  {http://adsabs.harvard.edu/abs/2013ApJ...765..100S} {765, 100}

\bibitem[\protect\citeauthoryear{{Tapping}}{{Tapping}}{2013}]{2013SpWea..11..394T}
{Tapping} K.~F.,  2013, \mn@doi [Space Weather] {10.1002/swe.20064}, \href
  {http://adsabs.harvard.edu/abs/2013SpWea..11..394T} {11, 394}

\bibitem[\protect\citeauthoryear{{Tapping} \& {Vald{\'e}s}}{{Tapping} \&
  {Vald{\'e}s}}{2011}]{2011SoPh..272..337T}
{Tapping} K.~F.,  {Vald{\'e}s} J.~J.,  2011, \mn@doi [\solphys]
  {10.1007/s11207-011-9827-1}, \href
  {http://adsabs.harvard.edu/abs/2011SoPh..272..337T} {272, 337}

\bibitem[\protect\citeauthoryear{{Tripathy}, {Jain}  \& {Hill}}{{Tripathy}
  et~al.}{2015}]{2015ApJ...812...20T}
{Tripathy} S.~C.,  {Jain} K.,   {Hill} F.,  2015, \mn@doi [\apj]
  {10.1088/0004-637X/812/1/20}, \href
  {http://adsabs.harvard.edu/abs/2015ApJ...812...20T} {812, 20}

\bibitem[\protect\citeauthoryear{{Upton} \& {Hathaway}}{{Upton} \&
  {Hathaway}}{2014}]{2014ApJ...780....5U}
{Upton} L.,  {Hathaway} D.~H.,  2014, \mn@doi [\apj]
  {10.1088/0004-637X/780/1/5}, \href
  {http://adsabs.harvard.edu/abs/2014ApJ...780....5U} {780, 5}

\bibitem[\protect\citeauthoryear{{Watson}, {Penn}  \& {Livingston}}{{Watson}
  et~al.}{2014}]{2014ApJ...787...22W}
{Watson} F.~T.,  {Penn} M.~J.,   {Livingston} W.,  2014, \mn@doi [\apj]
  {10.1088/0004-637X/787/1/22}, \href
  {http://adsabs.harvard.edu/abs/2014ApJ...787...22W} {787, 22}

\bibitem[\protect\citeauthoryear{{de Toma}, {Chapman}, {Preminger}  \&
  {Cookson}}{{de Toma} et~al.}{2013}]{2013ApJ...770...89D}
{de Toma} G.,  {Chapman} G.~A.,  {Preminger} D.~G.,   {Cookson} A.~M.,  2013,
  \mn@doi [\apj] {10.1088/0004-637X/770/2/89}, \href
  {http://adsabs.harvard.edu/abs/2013ApJ...770...89D} {770, 89}

\bibitem[\protect\citeauthoryear{{van Saders}, {Ceillier}, {Metcalfe}, {Silva
  Aguirre}, {Pinsonneault}, {Garc{\'{\i}}a}, {Mathur}  \& {Davies}}{{van
  Saders} et~al.}{2016}]{2016Natur.529..181V}
{van Saders} J.~L.,  {Ceillier} T.,  {Metcalfe} T.~S.,  {Silva Aguirre} V.,
  {Pinsonneault} M.~H.,  {Garc{\'{\i}}a} R.~A.,  {Mathur} S.,   {Davies} G.~R.,
   2016, \mn@doi [\nat] {10.1038/nature16168}, \href
  {http://adsabs.harvard.edu/abs/2016Natur.529..181V} {529, 181}

\makeatother
\end{thebibliography}

\end{document}